\documentclass[twocolumn,epjc3]{svjour3} 

\usepackage{amssymb}
\RequirePackage{graphicx}
\usepackage{newtxtext,newtxmath}

\usepackage{multirow}
\RequirePackage{amsmath}
\RequirePackage{flushend}

\RequirePackage[colorlinks,citecolor=blue,urlcolor=blue,linkcolor=blue]{hyperref}

 \newcommand{\ud}{\ensuremath{\mathrm{d}}}

\begin{document}

\title{Comment on ``Apparent horizons of the Thakurta spacetime and the description
of cosmological black holes"}

\author{Alan Maciel\thanksref{addr1}
        \and
        Vilson T. Zanchin\thanksref{addr2}
}

\institute{Centro de Matemática, Estatística e Ciências da Computação,  Universidade Federal do ABC, Avenida dos Estados 5001, 
09210-580 Santo Andr\'e, SP, Brazil\label{addr1}
            \and
Centro de Ci\^encias Naturais e Humanas,  Universidade Federal do ABC, Avenida dos Estados 5001, 
09210-580 Santo Andr\'e, SP, Brazil\label{addr2}
}

 \date{ }
\maketitle

\begin{abstract}
The findings discussed in "Apparent Horizons of the Thakurta Spacetime and the Description of Cosmological Black Holes" [Eur. Phys. J. C 82, 347 (2022] deviate from prior research, particularly our study in Phys. Rev. D 95, 084031 (2017). We clarify this inconsistency, which arises from employing a coordinate transformation covering a distinct region of spacetime—specifically one considered inaccessible due to its location beyond a singularity, making it unreachable from infinity. Additionally, we provide a concrete example where the distinct horizons and spacetime regions are readily discernible.
\end{abstract}

\section{Introduction}

Deciphering the dynamic solutions in General Relativity is a quite complex task. Each scenario demands detailed scrutiny, considering potential curvature singularities, infinities, and incomplete geodesics within the provided coordinates, and how they interplay with the inherent free functions of the solution. The McVittie metric \cite{McVittie:1933zz}, originating in 1933, serves as a prime example. Despite its age, achieving a comprehensive understanding of its intricacies has been the subject of numerous articles \cite{Nolan:1998xs,Nolan:1999kk,Nolan:1999wf,Kaloper:2010ec,Lake:2011ni,daSilva:2012nh} before a satisfactory elucidation was reached.

The Thakurta metric \cite{Thakurta1981} is similarly complex, even in its simplest spherically symmetric form without angular momentum. An in-depth analysis of this version of the Thakurta solution was presented in \cite{Mello:2016irl}, detailing potential singularities and incomplete geodesics based on the behavior of the scale factor $a(t)$. While $a(t)$ is mostly free, it must satisfy basic requirements such as $a(0) = 0$ and $\dot{a}(t) > 0$ to align with the expected properties of a scale factor in the standard cosmological model. In such a work, the conditions distinguishing each scenario were identified, complemented by causal diagrams where the apparent horizons were  plotted.

However, a recent claim in \cite{Kobakhidze:2021rsh} challenges previous findings, suggesting the existence of future-type apparent horizons in the Thakurta solution, contrary to the conclusions in \cite{Mello:2016irl}. In the spirit of scientific inquiry, it is crucial not only to publish diverse and sometimes conflicting results but also to investigate the underlying reasons for such alleged contradictions.
This article aims to explore these motives.

This article is structured as follows. In Section \ref{sec:Thakurta} we introduce the Thakurta metric and the main coordinate systems used in its study. Section \ref{sec:fakehorizon} presents a brief summary of the main findings of Ref.~\cite{Kobakhidze:2021rsh}. We analyze and identify the reasons for the alleged contradiction between the results in \cite{Kobakhidze:2021rsh} and in \cite{Mello:2016irl}in Section \ref{sec:explanation}. Finally, in Section \ref{sec:example} we provide a concrete example of Thakurta spacetime to illustrate the points discussed.

\section{The Thakurta metric} \label{sec:Thakurta}

The non-rotating Thakurta line element can be obtained by multiplying the spacelike components of the Schwarzschild line element, in Schwarzschild coordinates, by a time dependent scale factor function $a(t)$, what gives
\begin{gather} 
\ud s^2 = - f(r) \ud t^2 + a(t)^2\left( \frac{\ud r^2}{f(r)} + r^2 \ud \Omega^2\right), \label{eq:ThakurtaS}
\end{gather}
where $\ud \Omega^2= r^2 \ud \theta^2 + r^2 \sin^2 \theta\, \ud \phi^2$ and
\begin{equation}
f(r) = 1- \frac{2m}{r}, \label{eq:f(r)}
\end{equation}
with $m > 0$ being a constant that, for $a(t) =1$, corresponds to the ADM mass parameter.
For the present analysis, it is convenient to rewrite the Thakurta metric \eqref{eq:ThakurtaS} by using the areal radius $R=r\, a(t)$ as a coordinate, as presented in \cite{Mello:2016irl}, i.e.,
\begin{gather}
\ud s^2 = - F(t,R) \ud t^2 - \frac{2 H(t) R}{f(t,R)}\ud t \,\ud R + \frac{\ud R^2}{f(t,R)} + R^2 \ud \Omega^2 , \label{eq:Thakurtam}
\end{gather}
where 
\begin{align}
 & f(t,R) = 1 - \frac{2M(t)}{R}\,,\\ 
& F(t,R) =  f(t,R) - \dfrac{H(t)^2 R^2}{ f(t,R)},
\end{align}
with $M(t) = m\, a(t)$, $H(t) = \dfrac{\dot{a}(t)}{a(t)}$, and with the overdot $(\dot{~})$ standing for the derivative with respect to the time $t$.

\section{The future apparent horizon} \label{sec:fakehorizon}

We start by briefly reviewing the main result of \cite{Kobakhidze:2021rsh}.
From Eq.~\eqref{eq:Thakurtam}, two coordinate transformations were made in \cite{Kobakhidze:2021rsh}. The first transformation introduces the Kodama time $\ud\tau$, which, in spherical symmetry, corresponds to the time flow that is at each event tangent to the tube of spheres of constant areal radius $R$, or otherwise said, that is orthogonal to the 1-form $\ud R$ at each event. 
This has the advantage of making the metric diagonal, and brings the metric to a simple form when the spacetime is static.

 It is important to notice that the Kodama time coordinate always exists locally for spherically symmetric spacetimes \cite{Kodama:1979vn,Abreu:2010ru}, but such coordinates do not necessarily cover the maximum extension of the spacetime. One simple example is the Schwarzschild metric, where the Schwarzschild {standard} time coordinate corresponds to the Kodama time, but it cannot describe at the same time the spacetime outside and inside the event horizon.

In \cite{Kobakhidze:2021rsh}, the Kodama time was defined as
\begin{gather}
\ud \tau = e^{\phi(t,R)} \ud t + e^{\phi(t,R)} \frac{H(t)R}{f(t,R)} \frac{\ud R}{F(t,R)},
\end{gather}
where $\phi(t,R)$ must satisfy the integrability condition
\begin{gather}
\dfrac{\partial}{\partial R} \Big(e^{\phi(t,R)} \Big) = \dfrac{\partial}{\partial t} \left[ e^{\phi(t,R)} \frac{H(t)R}{f(t,R) F(t,R)} \right].
\end{gather}
Using the Kodama time, the Thakurta metric \eqref{eq:Thakurtam} is written in the form\footnote{Note that we are using the (- + + +) signature in this article, contrary to the (+ - - -) signature used in \cite{Kobakhidze:2021rsh}, see also \cite{Volovik}.}
\begin{gather}
\ud s^2 = -e^{-2\phi} F(\tau,R) \ud \tau^2 + \frac{\ud R^2}{F(\tau,R)} + R^2 \ud \Omega^2 . 
\end{gather}

Due to the coordinate singularities at $F(\tau,R) = 0$, which coincides with the apparent horizons, a new transformation of the time coordinate is needed, in order to find coordinates non-singular at this locus (or loci). The transformation proposed in \cite{Kobakhidze:2021rsh} is analogous to the one relating Painlevé-Gullstrand (PG) coordinates with Schwarzschild coordinates in Schwarzschild spacetime. The transformation is given by
\begin{gather}
\ud \tilde{\tau} = \alpha(\tau,R) \ud \tau + \alpha(\tau, R)\, e^{\phi(\tau,R)}\, \frac{\sqrt{1-F(\tau,R)}}{F(\tau,R)}\,\ud R\,, \label{eq:tildetau}
\end{gather}
where $\alpha(\tau, R)$ must satisfy the integrability condition
\begin{gather}
\dfrac{\partial\,\alpha(\tau,R)}{\partial R} = \dfrac{\partial}{\partial {\tau}}\left( \alpha(\tau, R)\, e^{\phi} \frac{\sqrt{1-F(\tau,R)}}{F(\tau,R)} \right)\,.\label{eq:intcond2}
\end{gather}
Then, the line element is written as
\begin{gather}
\ud s^2 =  - \frac{e^{-2\phi(\tilde\tau,R)}}{\alpha^2(\tilde\tau,R)} F(\tau,R) \ud \tilde{\tau}^2 \nonumber \\ \quad\;\; + 2 \frac{e^{-\phi(\tilde\tau,R)}}{\alpha(\tilde\tau,R)} \sqrt{1-F(\tilde\tau,R)}\,\ud \tilde{\tau} \ud R + \ud R^2 + R^2 \ud \Omega^2 \,.
\label{eq:falsePGmetric}
\end{gather}
The equation for the null rays ($\ud s^2 = 0$) gives us
\begin{gather}
\frac{\ud R}{\ud \tilde{\tau}} = \dfrac{e^{-\phi(\tilde\tau,R)}}{\alpha(\tilde\tau,R)}\left(\pm 1 - \sqrt{1-F(\tilde\tau,R)}\right),
\end{gather}
 where, in comparison with the notation in \cite{Kobakhidze:2021rsh}, we replaced $c$ with $e^{-\phi(\tilde\tau,R)}\big/\alpha(\tilde\tau,R)\,$.

Then, choosing the dual null basis ($l^\mu$,\, $k^\mu)$ as follows,
\begin{gather}
l^\mu = \left(\alpha\, e^\phi, 1 - \sqrt{1-F}, 0, 0 \right ), \\
k^\mu = \frac{1}{2}\left(\alpha\, e^\phi, -1 -\sqrt{1-F}, 0,0 \right),
\end{gather}
we can easily compute the respective null expansions $\Theta_l$ and $\Theta_k$,
\begin{gather}
\Theta_l = \frac{2}{R} \left(1- \sqrt{1-F}\right)  , \qquad
\Theta_k = -\frac{1}{R} \left(1 + \sqrt{1-F}\right)\,,
\end{gather}
which, for $F=0$, give $\Theta_l = 0$ and $\Theta_k < 0$. Therefore, the authors of Ref.~\cite{Kobakhidze:2021rsh} conclude that there is a future apparent horizon, what seems to be in disagreement with the findings in \cite{Mello:2016irl}.

\section{Examining the contradiction} \label{sec:explanation}

\subsection{On the integrability conditions}

Let us then look at the integrability condition \eqref{eq:intcond2}, 
\begin{align}
\alpha_R & =  \alpha_\tau e^{\phi} \frac{\sqrt{1-F}}{F} + \alpha \phi_\tau e^{\phi} \frac{\sqrt{1-F}}{F}\nonumber \\ &\;\; - \alpha e^{\phi} \sqrt{1-F}\frac{F_\tau}{F^2}  -  \alpha e^{\phi}\dfrac{1}{2F} \frac{F_\tau}{\sqrt{1-F}}, \label{eq:intcondition}
\end{align}
where we used the subscript notation for partial derivatives in order to shorten the expression.

The invariance under diffeomorphisms in general relativity means that the relationship between different coordinate systems must be $C^\infty$ in an open set of the spacetime described by the two coordinate systems. Therefore, in the spacetime patch described by both the Kodama coordinates and the PG-like coordinates, the derivatives of  $\alpha$ must be all finite, in particular $\alpha_R$. Since the dependence of $\alpha_R$ on $F$ and its derivatives is at least $\mathcal{O}\left(F^{-1}\right)$ in all terms, then, in the limit $F \to 0$, the partial derivatives $\alpha_\tau e^{\phi}$, $\alpha \phi_\tau e^{\phi}$, and the function $\alpha e^{\phi}$ must all vanish at least as fast as $\mathcal{O}(F^1)$.
These limiting situations can be met in two ways,
\begin{gather}
e^{\phi} \to 0,  \quad \text{with finite $\alpha$, $\alpha_\tau$, and $\phi_\tau$\,, or} \\
\alpha_\tau \to 0 \quad \text{and} \quad \alpha \to 0,\quad \text{with finite $e^{\phi}$}.
\end{gather} 
However, both conditions make the metric Eq.~\eqref{eq:falsePGmetric} ill defined at the limit $F \to 0$. This is consistent with the fact that the Kodama time coordinate is singular at the spacetime horizons, i.e., in the limit $F \to 0$. Therefore, we should understand \eqref{eq:intcondition} as implying that, in order for the Painlevé-Gullstrand coordinates not to be singular at $F \to 0$, the left hand side of Eq.~\eqref{eq:tildetau} should vanish, what implies that $\dfrac{\ud {\tau}}{\ud \tilde{\tau}}$  diverges. We interpret this divergence as a sign that as $F \to 0$ we have an infinite variation of $\tau$ corresponding to finite variation of $\tilde{\tau}$, meaning that the PG-like $\tilde{\tau}$ coordinate extends analytically the spacetime metric described by the Kodama coordinates.

We must note that in the analysis of the Thakurta spacetime made in \cite{Mello:2016irl} it was shown that, in cases where $\dot{a}(t) > 0$ for all times, there is a curvature singularity located at the surface $R = 2M(t)$. For this reason, only the region of the spacetime between the future infinity, the big-bang-like singularity at $a(t) = 0$, and the singularity at $R = 2M(t)$ was analyzed, and this region is completely described by the coordinates in Eq. \eqref{eq:Thakurtam}. In cases where $\dot{a}(t) = 0$ for $t > T$, for some finite $T$, there is no singularity at $R = 2M(t)$ and then an extension is possible and it was studied in \cite{Mello:2016irl}. The fact that the coordinates systems used do not describe exactly the same regions of the spacetime is one of the pieces of the apparent contradiction between \cite{Mello:2016irl} and \cite{Kobakhidze:2021rsh}.

\subsection{The role of the singularity}

In \cite{Kobakhidze:2021rsh}, advanced PG-like coordinates are chosen. One may also define retarded PG-like coordinates as 
\begin{equation}
\ud \tau_{\text{ret}} = \ud \alpha\, \ud \tau - \alpha e^\phi \frac{\sqrt{1-F}}{F} \ud R,
\end{equation}
in which case, the resulting line element would be
\begin{equation}
\ud s^2 = - \frac{e^{-2\phi}}{\alpha^2} F \ud \tilde{\tau}^2 -2 \frac{e^{-\phi}}{\alpha} \sqrt{1-F}\ud \tilde{\tau} \ud R + \ud R^2 + R^2 \ud \Omega^2.
\label{eq:falsePGmetric-retarded}
\end{equation}

In these coordinates, the null basis for \eqref{eq:falsePGmetric-retarded} given by
\begin{gather}
l^\mu = \big(\alpha\, e^\phi,\, 1 + \sqrt{1 - F},\, 0,\, 0\big), \nonumber\\
k^\mu = \frac12 \big(\alpha\, e^\phi ,\, -1 + \sqrt{1 - F},\, 0,\, 0\big),
\end{gather}
provides us with the expansions
\begin{gather}
\Theta_l = \frac{2}{R} (1 + \sqrt{1+F}), \qquad
\Theta_k = \frac{1}{R} (-1 + \sqrt{1-F}).
\end{gather}
Now we have, for $F \to 0$, $\Theta_k = 0$ and $\Theta_l > 0$, as in \cite{Mello:2016irl}, which means that the analyzed apparent horizons are not future horizons.

The explanation for the different results is that the horizons found with the advanced  PG-like coordinates and those found with the retarded  PG-like coordinates are different horizons, that are located at different loci, even though they correspond to the same equation $F = 0$ in both coordinates. In order to better understand this issue, we have to analyze more carefully all the solutions of the equation $F = 0$. Therefore, in order to evaluate the properties of the apparent horizons defined by this equation it is necessary to study each one and all of such solutions separately.

As mentioned above, one of the findings of \cite{Mello:2016irl} is that if $a(t)$ is unbounded as $t \to \infty$, there is a singularity at the locus $R = 2M(t)$. Therefore, the relevant  solutions of the equation $F(t,R) = 0$ for $R(t)$ are those satisfying the constraint $R(t) > 2M(t)$. Some solutions of $F = 0$ do not satisfy this criterion\footnote{{ The fact that the PG-Like coordinates employed in \cite{Kobakhidze:2021rsh} imply $f(r)<0$ was already noticed  in the work \cite{Harada:2021xze}, see also Ref.~\cite{Sato:2022yto}}.}. In fact, the equation
\begin{gather}
F = 1 - \frac{2M}{R} - \frac{H^2 R^2}{1 - \frac{2M}{R}} = 0,  
\end{gather}
provided $R \neq 2M$, unfolds into two second degree equations for $R(t) \neq 0$, namely,
\begin{gather}
\pm HR^2 +R - 2M= 0, \quad{\rm or}\quad \pm \dot a r^2 + r - 2m=0. \label{eq:2nddegree}
\end{gather}

By choosing the "-" sign in Eq.~\eqref{eq:2nddegree}, we find the two roots
\begin{gather}
R_{\pm} = \frac{1 \pm \sqrt{1-8HM}}{2H},\quad {\rm or}\quad  r_\pm = \frac{1 \pm \sqrt{1- 8m\dot a}}{2\dot a}. \label{eq:horizons1}
\end{gather}
\noindent
From the last relations for $R_{\pm}$, with $H> 0$ and $1> 8H\, M$, we have $R_{-} = \frac{1}{2H} \left(1 - \sqrt{1- 8HM} \right) > 2M$  and also $R_+ = \frac{1}{2H} \left(
1 + \sqrt{1-8HM} \right) > R_-$. This means that these two solutions correspond to horizons that are between the singularity $R = 2M$ and infinity for all times. This are the apparent horizons analyzed in \cite{Mello:2016irl}.

Considering now the "$+$" sign in Eq.\eqref{eq:2nddegree}, once again we obtain two solutions
\begin{gather}
R^{(+)}_\pm = \frac{-1 \pm \sqrt{1+ 8HM}}{2H}, \; {\rm or}\;\; r^{(+)}_\pm = \frac{-1 \pm \sqrt{1+ 8m\dot a}}{2\dot a}. \label{eq:horizons2}
\end{gather}
The reasonable assumptions $M > 0$ and $H > 0$ imply $R^{(+)}_- < 0$ for all times and, therefore, this is not a physical solution.  
For $R_+^{(+)}$, we obtain $
-1 + \sqrt{1+8HM} < -1 + 1 + 4HM  =4M\, H$, which implies $0< R^{(+)}_+ < 2M$. This means that $R^{(+)}_+(t)$ is behind the singularity $R=2M$. The properties of the apparent horizon $R_+^{(+)}(t)$ were not studied in \cite{Mello:2016irl}, since such a horizon is not connected to the physics at spacetime infinity. 

In sum, the solution found when we choose the advanced PG-like time is the innermost solution, which is a future apparent horizon, in agreement with what is stated in \cite{Kobakhidze:2021rsh}. This is consistent with the fact that the advance PG coordinates are suitable for future horizons.
However, as a more detailed analysis was not made in order to establish the actual location of those horizons, it was not acknowledged that a distinct patch of Thakurta spacetime was being analyzed, a patch that was only studied in \cite{Mello:2016irl} in the particular cases where the singularity at $R=2M$ was not present, and then the future trapping horizon was correctly identified.  These are particular cases that correspond to scale factors $a(t)$ that converges to a finite value $a_{\infty}$ as $t\to\infty$, in which cases $H$ vanishes fast enough towards time infinity, and the surface $R = 2M$ turns out to be an outer future trapping horizon, besides also being an event horizon.

It is also important to note that $t \to \infty$ along any causal curves as $R \to 2M$, such that the functions $a(t)$ and therefore $H(t)$ need to be extended beyond $t \to \infty$ in order to be possible to localize a horizon behind the surface $R = 2M$.

\section{The horizons of a CDM Thakurta model} \label{sec:example}

\begin{figure}[hb!]
\centering
\includegraphics[scale=1.]{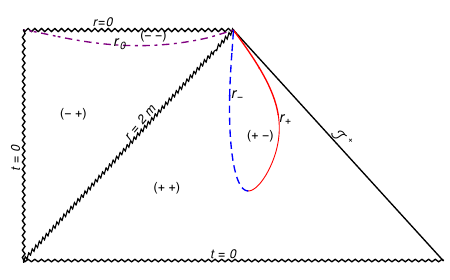} 
\caption{Conformal diagram describing two disconnected regions, the region between infinity and the $r=2m$ singularity, and region between the singularities $r-2m$ and $r=0$. The future apparent horizon is represented by the dot-dashed line labeled $r_0$.}
\label{fig:diagrama-dust}
\end{figure}

In order to provide a concrete example of what has been discussed throughout this article, we assume $a(t) = (t/t_0)^{2/3}$, as in a dust filled cosmological spacetime. The conformal diagram obtained is shown in Fig.~\ref{fig:diagrama-dust}.

Contrary to the procedure adopted in Ref.~\cite{Mello:2016irl}, here we are going to analyze also the patch of the spacetime behind the singularity, in order to show the presence of the future apparent horizon. 
This analysis is made clearer by making use of Kruskal-like coordinates, as in \cite{Mello:2016irl}, 
that can be more readily achieved by using the conformal time coordinate 
$$\displaystyle{\eta(t) = \int_0^{t} \frac{\ud \tilde{t}}{a(\tilde{t})}=3t_0^{2/3} t^{1/3}},$$
in terms of which, the line element acquires the form
\begin{gather}
\ud s^2 = a^2(\eta)\left( - f(r)\, \ud \eta^2 + \frac{\ud r^2}{f(r)}+  r^2 \ud \Omega^2\right), \label{eq:thakurtam2}
\end{gather}
with $f(r)$ given in Eq.~\eqref{eq:f(r)}.
For pedagogical reasons, we repeat some of the preceding analysis in these coordinates. 

We can then define the Kruskal-like coordinates for the region with $r > 2m$ as
\begin{subequations}
\begin{align}
T &=  \left| \frac{r}{2m}-1  \right| ^{1/2} e^{r/4m} \sinh \left( \frac{\eta}{4m} \right), \\
X &= \left| \frac{r}{2m}-1  \right| ^{1/2} e^{r/4m} \cosh \left( \frac{\eta}{4m} \right), 
\end{align} \label{eq:kruskalcoord}
\end{subequations} 
and  as
\begin{subequations}
\begin{align}
T &=  \left| \frac{r}{2m}-1  \right| ^{1/2} e^{r/4m} \cosh \left( \frac{\eta}{4m} \right), \\
X &= \left| \frac{r}{2m}-1  \right| ^{1/2} e^{r/4m} \sinh \left( \frac{\eta}{4m} \right), 
\end{align} \label{eq:kruskalcoord2}
\end{subequations} 
 for the region with $r < 2m$. These coordinates bring the line element \eqref{eq:thakurtam2} to the form
\begin{gather}
\ud s^2 =  \frac{32m^3 a^2}{r} e^{r/2m} \left( - \ud T^2 + \ud X^2 \right) + a^2 r^2 \ud \Omega^2.
\end{gather}

The null expansions based on these coordinates are obtained by taking the basis $k^{\mu}$ and $ l^{\mu}$ given by
\begin{gather}
k^{\mu} = \left(1/\sqrt{K},\, 1/\sqrt{K},\, 0,\, 0\right),\nonumber\\
l^{\mu} = \frac{1}{2}\left(1/\sqrt{K}, \,-1/\sqrt{K}, \,0,\, 0\right),
\end{gather}
where $K = \frac{32 m^3 a^2}{r} e^{r/2m} $, and that satisfy the constraint $k^\mu l_\mu = -1$. 
Hence, the result for the null expansions is 
\begin{gather}
\Theta_k = \frac{2}{a r\sqrt{K}}\left( \partial_T + \partial_X \right) (ar) , \nonumber \\
\Theta_l = \frac{1}{a r\sqrt{K}}\left( \partial_T - \partial_X \right) (ar) . \label{eq:expansion4}
\end{gather}

Now, by performing the inverse coordinate transformation $(T,\, X)$ $\to$  $(\eta,\, r)$, we get
\begin{gather}
\partial_T = \frac{4m}{X^2 - T^2} \left( X\partial_\eta - Tf(r)\partial_r \right), \nonumber\\
\partial_X = \frac{4m}{X^2 - T^2} \left( - T\partial_\eta + X f(r)\partial_r \right),
\end{gather}
\noindent
which provides us with
\begin{gather}
\partial_T + \partial_X = \frac{4m}{X + T}\left( \partial_{\eta} + f(r)\, \partial_r \right), \nonumber\\
\partial_T - \partial_X = \frac{4m}{X-T} \left( \partial_{\eta} - f(r)\, \partial_r \right).
\end{gather}
\noindent
The null expansions from eq.~\eqref{eq:expansion4} then read
\begin{equation}
\begin{aligned}
\Theta_k = \frac{8m}{r \sqrt{K} (X + T) } \left( \frac{a'(\eta)}{a(\eta)}r + f(r) \right),\\
\Theta_l = \frac{4m}{r \sqrt{K} (X -T) } \left( \frac{a'(\eta)}{a(\eta)}r - f(r) \right).  \label{eq:expansions5}
\end{aligned}
\end{equation}

The apparent horizons correspond to the zeros of one of the null expansions, implying $\Theta_k\Theta_l= 0$. By using Eqs.~\eqref{eq:expansions5}, it is seen that the zeros of such a product leads to 
\begin{equation}
    \frac{a'^2(\eta)}{a^2(\eta)}r^2- f^2(r) = \dot a^2(t)\, r^2 -f^2(r) =0, \label{eq:trapping1}
\end{equation}
which is the same as Eq.~\eqref{eq:2nddegree}. Naturally, this equation leads to the two situations discussed above, as we comment also in the following

By assuming that $\dot{a} > 0$ and, moreover, $r > 2m $ it follows $f(r) > 0$. In such a case, the apparent horizons correspond to the situation
\begin{gather}
\Theta_k > 0\,,\qquad 
\Theta_l = 0\,,
\end{gather}
which characterizes past horizons, as expected from our discussion in Sec.~\ref{sec:explanation}. In fact, these conditions imply the equation 
\begin{equation}
   \dot{a}\, r^2 - r + 2m = 0, 
\end{equation}
whose solutions $r_\pm(t)$ are given in Eq.~\eqref{eq:horizons1}. The horizons  $r_\pm(t)$ appear in the conformal diagram of Fig.~\ref{fig:diagrama-dust} as the red solid line labeled by $r_+$, and as the blue dashed line labeled by $r_-$, respectively.

On the other hand, the conditions $\dot{a} > 0$ and $0< r < 2m $ imply that $f(r)<0$, and in this case the apparent horizons correspond to
\begin{gather}
\Theta_k = 0\,, \qquad \Theta_l < 0\,,
\end{gather}
which characterizes future horizons, as found in \cite{Kobakhidze:2021rsh}. In fact, these conditions imply the equation 
\begin{equation}
   \dot{a}\, r^2 + r - 2m = 0, 
\end{equation}
whose solutions $r^{(+)}_\pm(t)$ are given in Eq.~\eqref{eq:horizons2}.  The solution $r^{(+)}_-(t)$ is not a horizon, since it assumes negative values. In turn, the solution 
\begin{equation}
r_0(t)\equiv r^{(+)}_+(t)= \frac{-1 + \sqrt{1+ 8m\dot a}}{2\dot a}
\end{equation}
is the future apparent that appears in the conformal diagram of Fig.~\ref{fig:diagrama-dust} as the purple dash-dotted line labeled by $r_0$.  

We may also verify that the horizon $r_0(t)$ is indeed spacelike as it appears in Fig.~\ref{fig:diagrama-dust}. Such a horizon is behind the singularity, i.e., it satisfies  $f(r_0)< 0$, and gives
\begin{gather}
1 - \frac{2m}{r_0(t)} + \dot{a}r_0(t) = 0. \label{eq:fakehorizon}
\end{gather}
From Eq.~\eqref{eq:fakehorizon}, we obtain the 1-form normal to the horizon,
\begin{gather}
n_\mu =\left(\ddot{a}\, r_0(t),\, \frac{2m}{r_0^2(t)} + \dot{a},\, 0 ,\, 0 \right),
\end{gather}
\noindent
whose norm is given by
\begin{gather}
||n||^2 = \frac{\ddot{a}^2\,r_(t)}{\dot{a}} \left(1 - \frac{\dot{a}^2}{a^2\ddot{a}^2 r_0^2(t)}\left[\frac{4m}{r_0(t)} -1\right]^2\right),
\end{gather}
which, for the dust universe model, $a(t) = \left(t/t_0\right)^{2/3}$, is negative for all times, implying that the horizon is spacelike in its entirety.

\begin{figure}[t!]
\centering
\includegraphics[width=0.4\textwidth]{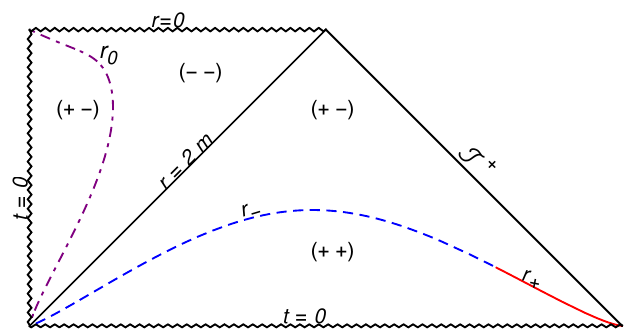} 
\caption{Conformal diagram describing a black hole for $a(t)= \tanh^{2/3}(t/t_0)$, $t>0$. The physical region between infinity and the $r=2m$ event horizon, and the trapped region between the horizon $r=2m$ and the singularities $r=0$ and $t=0$ (vertical). The future apparent horizon is represented by the dot-dashed line labeled $r_0$.}
\label{fig:diagrama-tanh23}
\end{figure}
  
In order to illustrate the behavior of the future apparent horizon in a case where the lightlike surface $r=2m$ is not a singularity, we analise the case with $a(t) = \tanh^{2/3}(t/t_0)$, in which case the scale factor $a(t)$ tends to a constant as $t\to\infty$, modeling an initial coupling with a dust filled universe followed by the decoupling at a finite time scale.  The conformal diagram for this scale factor is shown in Fig.~\ref{fig:diagrama-tanh23}.
In this case, the horizon $r_0(t)$ is spacelike for small $t$, then becomes timelike and tends to a lightlike surface as $t \to \infty$.

\section{Conclusion}

The findings of Ref.\cite{Kobakhidze:2021rsh} are all sustained on a Painlevé-Gullstrand-like transformation that entails an analytical continuation of the coordinate patch compared to the so-called Kodama coordinates. Interpreting results within such a coordinate system demands careful analysis to discern potential curvature singularities alongside coordinate singularities, which is pivotal for determining the physical significance of the analytical continuation. The absence of these procedures in Ref.~\cite{Kobakhidze:2021rsh} led to erroneous conclusions about the nature of apparent horizons for different observers. It is worth noticing that the nature of horizons remains consistent across observers or coordinate systems, as they are determined by the evaluation of spacetime scalars. What varies is the appearance or absence of specific loci in each coordinate system, what can lead to errors when comparing different loci in distinct coordinate systems as if they were identical. In the end, our analysis here shows that there is no contradiction between the findings in \cite{Kobakhidze:2021rsh} and \cite{Mello:2016irl}.


\begin{thebibliography}{10}
\providecommand{\url}[1]{{#1}}
\providecommand{\urlprefix}{URL }
\expandafter\ifx\csname urlstyle\endcsname\relax
  \providecommand{\doi}[1]{DOI \discretionary{}{}{}#1}\else
  \providecommand{\doi}{DOI \discretionary{}{}{}\begingroup \urlstyle{rm}\Url}\fi

\bibitem{McVittie:1933zz}
G. C. McVittie, The mass-particle in an expanding universe, Mont. Not. R.\ Astron.\ Soc. \textbf{93}, 325 (1933).
\newblock \doi{10.1093/mnras/93.5.325}

\bibitem{Nolan:1998xs}
B. C. Nolan, A point mass in an isotropic universe: Existence, uniqueness, and basic properties, Phys. Rev. D \textbf{58}, 064006 (1998).
\newblock \doi{10.1103/PhysRevD.58.064006}

\bibitem{Nolan:1999kk}
B. C. Nolan, A point mass in an isotropic universe: {{II. Global}} properties, Classical Quantum Gravity \textbf{16}, 1227 (1999).
\newblock \doi{10.1088/0264-9381/16/4/012}

\bibitem{Nolan:1999wf}
B. C. Nolan, A point mass in an isotropic universe: {III. The} region $r \leq 2 m$, Classical Quantum Gravity  \textbf{16}, 3183 (1999).
\newblock \doi{10.1088/0264-9381/16/10/310}

\bibitem{Kaloper:2010ec}
N.~Kaloper, M.~Kleban, D.~Martin, {McVittie's} legacy: Black holes in an expanding universe, Phys. Rev. D \textbf{81}, 104044 (2010).
\newblock \doi{10.1103/PhysRevD.81.104044}

\bibitem{Lake:2011ni}
K.~Lake, M.~Abdelqader, More on {McVittie's} legacy: A {Schwarzschild--de~Sitter} black and white hole embedded in an asymptotically {$\Lambda$CDM} cosmology, Phys. Rev. D \textbf{84}, 044045 (2011).
\newblock \doi{10.1103/PhysRevD.84.044045}

\bibitem{daSilva:2012nh}
A. M. da~Silva, M.~Fontanini, D. C. Guariento, How the expansion of the universe determines the causal structure of {McVittie} spacetimes, Phys. Rev. D \textbf{87}, 064030 (2013).
\newblock \doi{10.1103/PhysRevD.87.064030}

\bibitem{Thakurta1981}
S. N. G. Thakurta, Kerr metric in an expanding universe, Indian J. Phys. \textbf{55B}, 304 (1981).
\newblock \doi{10.1103/PhysRevD.81.064001}

\bibitem{Mello:2016irl}
M. M. C. Mello, A.~Maciel, V. T. Zanchin, {Evolving black holes from conformal transformations of static solutions}, Phys. Rev. D \textbf{95}, 084031 (2017).
\newblock \doi{10.1103/PhysRevD.95.084031}

\bibitem{Kobakhidze:2021rsh}
A.~Kobakhidze, Z. S. C. Picker, {Apparent horizons of the Thakurta spacetime and the description of cosmological black holes}, Eur. Phys. J. C \textbf{82}, 347 (2022).
\newblock \doi{10.1140/epjc/s10052-022-10312-1}

\bibitem{Kodama:1979vn}
H.~Kodama, Conserved energy flux for the spherically symmetric system and the back reaction problem in the black hole evaporation, Prog.\ Theor.\ Phys. \textbf{63}, 1217 (1980).
\newblock \doi{10.1143/PTP.63.1217}

\bibitem{Abreu:2010ru}
G.~Abreu, M.~Visser, Kodama time: Geometrically preferred foliations of spherically symmetric spacetimes, Phys. Rev. D \textbf{82}, 044027 (2010).
\newblock \doi{10.1103/PhysRevD.82.044027}

\bibitem{Volovik}  G. E. Volovik, 
Macroscopic quantum tunneling: From quantum vortices to black holes and Universe
    J. Exp. Theor. Phys. {\bf 135}, 388 (2022).

\bibitem{Harada:2021xze}
T.~Harada, H.~Maeda, and T.~Sato,
Thakurta metric does not describe a cosmological black hole,
Phys. Lett. B \textbf{833}, 137332 (2022).
\doi{10.1016/j.physletb.2022.137332}


\bibitem{Sato:2022yto}
T.~Sato, H.~Maeda and T.~Harada,
Conformally Schwarzschild cosmological black holes, 
Classical Quantum Gravity \textbf{39}, 215011 (2022)
[erratum: Classical Quantum Gravity  \textbf{40}, 079501 (2023)].
\doi{10.1088/1361-6382/ac902f}


\end{thebibliography}
\end{document}